\documentclass[review,5p,times,twocolumn,sort&compress]{elsarticle}

\usepackage{amssymb}

\usepackage[utf8]{inputenc}
\usepackage[usenames]{color}
\usepackage{array}
\usepackage{amsmath}
\usepackage{graphicx}
\usepackage[colorlinks=true] {hyperref}
\usepackage{xspace}
\usepackage{tabularx}
\usepackage{subfig}
\usepackage{color}
\usepackage{tabularx}

\journal{Materials Science in Semiconductor Processing}

\newcommand{\E}[2]{#1\times10^{#2}}
\newcommand{\Unit}[2]{ {#1}^{(\mathrm{ #2})}  }

\newcommand{\Atlas}{Atlas$^{\text{\textregistered}}$\xspace}
\newcommand{\Silvaco}{Silvaco$^{\text{\textregistered}}$\xspace}

\newcommand{\X}{{x}}
\newcommand{\WF}{W_f}
\newcommand{\N}{N_d}
\newcommand{\ThickInGaN}{T_{\mathrm{InGaN}}}

\begin{document}

\begin{frontmatter}



\title{Numerical Simulation of InGaN Schottky Solar Cell}

\author[afful,affcs]{Sidi Ould Saad Hamady\corref{cor1}}
\ead{sidi.hamady@univ-lorraine.fr}

\author[afful,affcs]{Abdoulwahab Adaine}
\ead{abdoulwahab.adaine@univ-lorraine.fr}

\author[afful,affcs]{Nicolas Fressengeas}
\ead{nicolas@fressengeas.net}

\cortext[cor1]{Corresponding author}

\newcommand{\LMOPS}{Laboratoire Matériaux Optiques, Photonique et Systèmes, EA 4423, Metz, F-57070, France}

\address[afful]{Universit\'e de Lorraine, \LMOPS}

\address[affcs]{CentraleSup\'elec, \LMOPS}

\begin{abstract}
The Indium Gallium Nitride (InGaN) III-Nitride ternary alloy has the potentiality to allow achieving high efficiency solar cells through the tuning of its band gap  by changing the Indium composition. It also counts among its advantages a relatively low effective mass, high carriers’ mobility, a high absorption coefficient along with good radiation tolerance.
However, the  main drawback of InGaN is linked to its p-type doping, which is difficult to grow in good quality and on which ohmic contacts are difficult to realize. The Schottky solar cell is  a good alternative to avoid the p-type doping of InGaN.
 In this report, a comprehensive numerical simulation, using mathematically  rigorous optimization approach based on state-of-the-art optimization algorithms, is used to find the optimum geometrical and physical parameters that yield the best efficiency of a Schottky solar cell within the achievable device fabrication range. A $18.2\%$ efficiency is predicted for this new InGaN solar cell design.

\end{abstract}


\end{frontmatter}



\section{Introduction}
\label{introduction}

The Indium Gallium Nitride (InGaN) III-Nitride ternary alloy can lead to high efficiency solar cells. Indeed its band gap can cover the whole solar spectrum, including the visible region, solely by changing the Indium composition \cite{bechstedt_energy_2003,reichertz_demonstration_2009,fabien_guidelines_2014}. The InGaN alloy also counts among its advantages relatively low effective mass and high mobilities for electrons and holes \cite{hsu_modeling_2008}, a high absorption coefficient \cite{wu_band_2003,polyakov_low-field_2006,jani_design_2007}  as well as a good radiation tolerance \cite{wu_superior_2003}, allowing its operation in extreme conditions.

However, the main drawbacks are the poor InGaN crystal quality, the difficulty to grow InGaN with Indium content covering the interesting range for solar application \cite{yamamoto_metal-organic_2013}, the difficulty of p-type doping mainly due to the high residual donors’ concentration and the lack of \emph{ad. hoc.} acceptors \cite{dahal_ingan/gan_2009}, and the difficulty to realize ohmic contacts on p-doped layers  \cite{jani_development_2008}. For these reasons the InGaN based solar cell is still in early development stages and the reported photovoltaic efficiency is still very low to be competitive with other well established III-V and silicon technologies  \cite{toledo_ingan_2012}. It is then vital to leverage these drawbacks and to develop alternative technologies. One possible paths could be the Schottky technology. This technology is largely used elsewhere in the III-Nitride based power devices and photodetectors \cite{saito_group_2009} but is very new to the InGaN photovoltaic technology. The first experimental work was published in 2009 by Xue Jun-Jun  \cite{jun-jun_au/pt/ingan/gan_2009}. The prototype developed by this team demonstrates the feasibility of such a concept: an optimization work becomes necessary in order to evaluate precisely and realistically the potentialities of this device. Some work has recently been done in that direction \cite{mahala_metal/ingan_2015}, using simplifying assumptions in order to pursue an analytical approach.

That is the reason why we propose, in this report and for the first time using a comprehensive numerical method, a detailed study of the potentialities of the InGaN Schottky technology as a viable alternative to the InGaN p-n junction solar cell.
We used mathematical algorithms to study this new solar cell where usual approaches use one-by-one parametric analysis which is inherently inexact and quite tedious. Our mathematical optimization approach is novel in the area of solar cell devices, though relatively common in other applied physics areas such as mechanical engineering.

The following section describes the physical modeling of the InGaN Schottky solar cell structure and discusses its main adjustable parameters. The second section presents the detailed optimized results and a device analysis with respect to the main cell parameters.

\section{Schottky Solar Cell Modeling}

\subsection{Physical Modeling}

\begin{figure}
  \includegraphics[width=\linewidth]{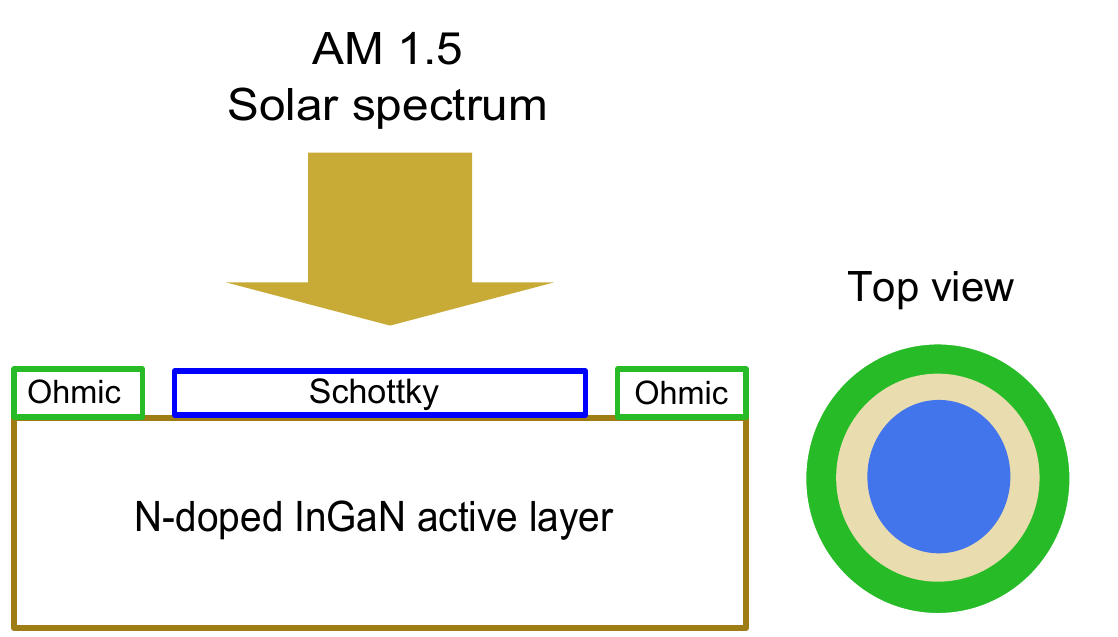}
  \caption{InGaN solar cell device structure.}
  \label{SchottkInGaN_rev_Fig1}
\end{figure}

The InGaN Schottky solar cell schematic view is shown on Figure \ref{SchottkInGaN_rev_Fig1}.

The here proposed device could be realized in practice using the conventional growth of InGaN on insulating substrates such as sapphire \cite{matsuoka_progress_2005} or Fe compensated GaN. To further ensure practical realization, we propose a coplanar device metalization.

 Modeling this device implies taking into account the interactions of the basic transport equations, including the Poisson and continuity equations on electrons and holes. In this model, the current densities need to be calculated including both the drift and diffusion components.

Two of the main key parameters are the mobilities of electrons and holes. Their values can be deduced from temperature and doping using the Caughey-Thomas expressions \cite{schwierz_electron_2005}:
\newcommand{\TRatio}{\left(\frac{T}{300}\right)}
\begin{equation}
\label{eq_mobility}
  \mu_m=\mu_{1m}\TRatio^{\alpha_m}+\frac{\mu_{2m}\TRatio^{\beta_m}- \mu_{1m}\TRatio^{\alpha_m}}{1+\left(\frac{N}{N_m^{\mathrm{crit}}\TRatio^{\gamma_m}}\right)^{\delta_m}}
,
\end{equation}
 where $m$ is either $n$ or $p$, $\mu_n$ being the electrons mobility and $\mu_p$ that of holes. $T$ is the absolute temperature. $N$ is the N-dopant (usually Silicon) density. $N^{\mathrm{crit}}$ and the $n$ or $p$ subscripted $\alpha$, $\beta$, $\delta$ and $\gamma$ are the model parameters which depend on the Indium composition \cite{farahmand_monte_2001}.

In addition to the mobility model, we took into account the bandgap narrowing effect \cite{lee_band-gap_1999}, the Shockley–Read–Hall (SRH) \cite{brandt_recombination_1998} and direct and Auger recombination models using the Fermi statistics \cite{delaney_auger_2009}.

Finally, the holes and electrons life time was taken equal to 1ns \cite{kumakura_minority_2005} in GaN, InN and InGaN.

\subsection{Material parameters}

All these models include a number of material dependent parameters which can be determined for a given material either from experimental work or \emph{ab initio} calculations. Their values are of course crucial for the numerical calculations to be meaningful, so as to give a good insight into the underlying physics and correctly pave the way to the realization of actual devices.
Fortunately, many of the parameters have been intensively studied in the past \cite{farahmand_monte_2001,davydov_band_2002,polyakov_low-field_2006,wu_band_2003} for GaN and InN binaries. Their values are summarized in Table \ref{tab_param}.

\begin{table}

\begin{center}
\subfloat[ Experimental data  from refs \cite{farahmand_monte_2001,davydov_band_2002,polyakov_low-field_2006,wu_band_2003}]{
\begin{tabular}{|c|c|c|c|c|c|}
\hline
~ & $\Unit{E_g}{eV}$ & $\Unit{\chi}{eV}$ & $\Unit{N_c}{cm^{-3}}$ & $\Unit{N_v}{cm^{-3}}$ & $\varepsilon$ \\ \hline
GaN & 3.42 & 4.1 & $\E{2.3}{18}$ & $\E{4.6}{19}$ & 8.9\\ \hline
InN & 0.7 & 5.6 & $\E{9.1}{17}$ & $\E{5.3}{19}$ & 15.3  \\ \hline
    \end{tabular}
\label{tab_physparam}
} 

\subfloat[Experimental data from refs \cite{polyakov_low-field_2006,nawaz_tcad-based_2012} except for $\alpha_n$, $\beta_n$ and $\gamma_n$ which have been estimated to 1 in the absence of any experimental data.]{\begin{tabular}{|c|c|c|c|c|c|c|c|}
\hline
~ &  $\Unit{\mu_n^1}{cm^2/Vs}$ &  $\Unit{\mu_n^2}{cm^{2}/Vs}$   & $\delta_n$ & $\Unit{N_n^{\mathrm{crit}}}{cm^{-3}}$ \\ \hline
GaN & 295  & 1460  & 0.71 & $\E{7.7}{16}$\\ \hline
InN & 1030  & 14150  & 0.6959 & $\E{2.07}{16}$ \\ \hline
    \end{tabular}
\label{tab_mobilparam_n}}

\subfloat[Experimental data from ref \cite{brown_finite_2010} except for $\alpha_p$, $\beta_p$ and $\gamma_p$ which have been estimated to 1 in the absence of any experimental data.]{\begin{tabular}{|c|c|c|c|c|c|c|c|}
\hline
~ &  $\Unit{\mu_p^1}{cm^2/Vs}$ &  $\Unit{\mu_p^2}{cm^{2}/Vs}$   & $\delta_p$ & $\Unit{N_p^{\mathrm{crit}}}{cm^{-3}}$ \\ \hline
GaN & 3  & 170  & 2 & $\E{1}{18}$\\ \hline
InN & 3  & 340  & 2 & $\E{8}{17}$ \\ \hline
    \end{tabular}
\label{tab_mobilparam_p}}
\end{center}
\caption{InN and GaN physical model parameters. Table \ref{tab_physparam} shows  the band-gap $Eg$, the electron affinity $\chi$,   the effective density of states $N_v$ and $N_c$ in the valence and conduction band respectively and the dielectric permittivity  $\varepsilon$. Table \ref{tab_mobilparam_n} and \ref{tab_mobilparam_p} summarize the parameters used to model the electron and hole mobilities in equation \eqref{eq_mobility}.}
\label{tab_param}
\end{table}

As we will treat the $x$ Indium composition in $\mathrm{In_xGa_{1-x}N}$ as a free parameter for the optimization that will be undertook in section \ref{optimparameters}, we will need the material parameters, as in Table \ref{tab_param}, for all Indium composition: $\forall x\in\left[0,1\right]$.

\begin{table}
  \begin{center}
\begin{tabular}{|l|c|c|}
\hline
Indium Composition & $\Unit{C}{eV^{-1}}$ & $\Unit{D}{eV^{-2}}$ \\ \hline
1	&0.69642	&0.46055\\\hline
0.83	&0.66796	&0.68886\\\hline
0.69	&0.58108	&0.66902\\\hline
0.57	&0.60946	&0.62182\\\hline
0.5	&0.51672	&0.46836\\\hline
0	&3.52517	&-0.65710\\\hline
  \end{tabular}
  \end{center}
\caption{Values for $C$ and $D$ in equation (\ref{eq_absorption}) as found by Brown \emph{et. al.} in \cite{brown_finite_2010}.}
\label{tab_Brown}
\end{table}

These valued are yielded by the standard \emph{modified Vegard law}. For the electronic affinity and the band-gap, the quadratic bowing factor is conventionally taken to be $b=1.43 \mathrm{eV}$ \cite{wu_band_2003}. However, some uncertainty remains concerning the actual value of $b$, as is discussed in section \ref{bowing}. For all the others parameters, the bowing factor is assumed null.

 In short, and provided the bowing factor is only taken in account for the band-gap energy and the electronic affinity, if $A^{\mathrm{{InN}}}$ and $A^{\mathrm{GaN}}$ are any one of the parameters in Table \ref{tab_param}, the corresponding parameter for $\mathrm{In_xGa_{1-x}N}$ will be deduced by :
\begin{equation}
A^{\mathrm{In_xGa_{1-x}N}}=xA^{\mathrm{InN}}+\left(1-x\right)A^{\mathrm{GaN}}-bx\left(1-x\right)
.
\label{eq_Vegard}
\end{equation}

The electron and holes mobility must be paid here a special attention:  the linear interpolation of the parameters included in equation \eqref{eq_mobility} implies that the mobility itself is \emph{not} interpolated in a linear way.

Modeling the Schottky solar cell implies also the need for a detailed model of light absorption in the whole solar spectrum and for all $x$ Indium composition. We propose to rely on a phenomenological model proposed previously \cite{brown_finite_2010} as

\begin{equation}
  \Unit{\alpha}{cm^{-1}}=\Unit{10^5}{cm^{-1}}\sqrt{C\left(E_{ph}-E_g\right)+D\left(E_{ph}-E_g\right)^2},
  \label{eq_absorption}
\end{equation}

where $E_{ph}$ is the incoming photon energy and $E_g$ is the material band gap at a given Indium composition.
Once $C$ and $D$ are known for a given Indium composition, the above equation \eqref{eq_absorption} yields the absorption coefficient for the whole solar spectrum.

The values for $C$ and $D$ are again taken from the experimental measurement reported in \cite{brown_finite_2010} and summarized in Table \ref{tab_Brown}. Their dependency on the Indium composition $\X$ is approximated by a polynomial fit, of the $\mathrm{4}^{th}$ degree for the former, and quadratic for the latter :

\begin{eqnarray*}
C &=& 3.525 - 18.29\X + 40.22\X^2 - 37.52\X^3 + 12.77\X^4\\
D &=& -0.6651 + 3.616\X - 2.460\X^2
\end{eqnarray*}

For the refraction index we used the Adachi model \cite{djurisic_modeling_1999}, defined for a given photon energy as:


\newcommand{\ERatio}{\frac{E_{ph}}{E_g}}
\begin{equation}
  n\left(E_{ph}\right)=\sqrt{\frac{A}{\left(\ERatio\right)^2}\left[2-\sqrt{1+\ERatio}-\sqrt{1-\ERatio}\right]+B },
\end{equation}


with $A$ and $B$ experimentally measured \cite{nawaz_tcad-based_2012,brown_finite_2010} for GaN ($A^{\mathrm{GaN}}=13.55$ and $B^{\mathrm{GaN}}=2.05$) and InN ($B^{\mathrm{InN}}=2.05$ and $B ^{\mathrm{InN}}=3.03$) and linearly interpolated for InGaN.

\par{Finally, we chose not to include the spontaneous polarization effect in our Schottky one-active-layer design as we expect its impact to be more sensitive in the more complex designs such as GaN/InGaN multijunction solar cells where it can negatively impact the solar cell efficiency or, on the contrary, enhance the efficiency \cite{lestrade2011modeling, thosar2013modeling}.}

\subsection{Optimization parameters}
\label{optimparameters}

The device is simulated in the framework of a drift-diffusion model using the \Atlas device simulation framework from the commercial \Silvaco software suite.
These pieces of software solve the above described non linear partial differential problem using the Newton-Raphson direct and the Gummel iterative methods \cite{kramer_semiconductor_1997}. These equations are solved in a two-dimensional framework.

For any given set of physical and geometrical parameters, \Atlas solves the model equations and outputs a whole set of data, including the current-voltage curve, the photocurrent spectral response, the electrical field distribution, the recombination variation, and the photovoltaic parameters among which can be found the short-circuit current, the open-circuit voltage, the fill factor and the conversion efficiency.

Doing a careful optimization of the above described solar cell implies to define precisely the optimization goal and the parameters that will be set as free and looked for, leaving the others as simple data to the problem. This is a strategic choice because it will deeply impact the optimization method and strategy able to solve the problem.

On the one hand, the choice of the figure of merit that will be maximized is crucial, as it will be the one and only factor that will be maximized. Its quantitative evaluation thus needs to be as quick and precise as can be. We chose here to maximize the conversion efficiency of the cell, as it can be evaluated from the data provided by \Atlas.

On the other hand, setting a large number of parameters as free surely increases the chances of getting the best merit factor possible, but as surely increases the problem difficulty and thus the computing time.

In the above described Schottky solar cell, we need to identify the two sets of relevant physical and geometrical parameters, that can be tuned and which have a significant impact on the cell behavior. In these two sets, we will strategically chose the parameters we will look for and those which are to be set as constant data.

In the geometrical set are to be found the cell width, length and height. The latter is the sum of the InGaN layer thickness and that of the metal electrodes used for the Schottky contact. The width and length of the cell define the cell area. We do here the simplifying assumptions that the results we will get are independent of the cell area, at least as far as the conversion efficiency is concerned. This area is thus arbitrarily set to a $\mathrm{1\mu m^2}$ square.

We have chosen to set the metal layer thickness to $5\mathrm{nm}$ and included in the simulation the work function as a variable parameter. Indeed, once the metal is chosen according to its work function, the optimization of the metal layer mainly requires geometrical considerations, such as interdigited electrodes, which is not within the scope of this analysis.

The InGaN absorbing layer thickness plays a significant role and we have chosen to let it be a free parameter for which we seek the optimum value.

The InGanN layer physical parameters that can be tuned and that have a significant impact on the cell behavior include its n-type doping and its Indium composition. We have chosen to let both of them be free parameters in order to find their optimum values.

Summarizing, the four free parameters are thus the Indium composition and the metal work function, along with the doping and the thickness of the n-type layer.

Finally, we have chosen to shine on the cell the ASTM-G75-03 solar spectrum taken from the National Renewable Energy Laboratory database \footnote{\url{http://rredc.nrel.gov/solar/spectra/am1.5/astmg173/astmg173.html}}.

\subsection{Parameters ranges and normalization}

As described above, the optimization task we have now to undertake is the optimization of the cell efficiency $\eta$, which is a function of four parameters, the InGaN layer thickness, the n-type doping, the Indium composition and the metal work function.

These parameters must however be constrained within a variation domain, which is the interval in which they are meaningful in physical and technological terms. The Indium composition is $\X$ in $\mathrm{In_xGa_{1-x}N}$ and will vary in the whole range: $\X\in\left[0, 1\right]$.

The metal work function $\WF$ is to be constrained roughly between the minimum and maximum work functions of existing and usable metals and alloys. We chose to constrain it between $5.1$ and $6.3 \mathrm{eV}$ : $\WF\in\left[5.1, 6.3\right]$.

The InGaN layer n-type doping $\N$ was chosen between the minimal residual N doping of $10^{16}\mathrm{cm^{-3}}$. Its maximum value was, on its side, a large $10^{19}\mathrm{cm^{-3}}$ if experimental data are considered. We get $\N\in\left[10^{16}, 10^{19}\right]\mathrm{cm^{-3}}$.

The last parameter we have to set bounds for is the the InGaN layer thickness $\ThickInGaN$. We chose to set its minimum value to a $100\mathrm{nm}$ thickness which ensures that a significant part of the wavelengths included in the solar spectrum are absorbed \cite{brown_finite_2010}. Its maximum was set to a large $1\mathrm{\mu m}$, obtaining $\ThickInGaN\in\left[0.1, 1\right]\mathrm{\mu m}$.

Finally, the state-of-the-art optimizer will treat the parameters as dimensionless quantities. Therefore, in order for the general optimizing algorithm to work as expected, all the parameters should have similar variation range. This is easily obtained for $\X$, $\WF$ and $\ThickInGaN$ by normalizing them to their maximum, and using this normalized values as the actual parameters.

The case of $\N$ is trickier. If we applied the same procedure, the algorithm would focus on the highest dopings, namely the last decade, which is something we want to avoid. The actually meaningful quantity is the $\log\left(\N\right)$. We chose the actual parameter for the n-type doping to be $\log_{10}\left(\N\right)$ and normalized it to its maximum value of $19$.

For the sake of clarity, the results presented in the following will be un-normalized, though the actual computation has indeed been done with the normalized parameters.

\subsection{Optimization}

The mathematical problem we must now solve is a bounded non linear optimization problem. In other words, we need to find the optimum quadruplet $\left(\X,\WF,\N,\ThickInGaN\right)$ within the four dimension hypercube \[\left[0,1\right]\times\left[5.1,6.3\right]eV\times[10^{16},10^{19}]\mathrm{cm^{-3}}\times\left[0.1,1\right]\mathrm{\mu m}\] which allows the maximum efficiency of the cell, computed in the way defined above.

The problem is non-linear, as the efficiency is not a linear function of all the 4 parameters we use here. It is also a problem for which the gradient of the efficiency with respect to the 4 parameters cannot be evaluated analytically, owing to the complexity of the efficiency computation.
We therefore need to use bounded optimization methods that do not rely on the optimized parameter gradient. 

Furthermore, the time needed to compute a single value of the efficiency can be counted from minutes to tens of minutes and the overall time can vary from a few hours to a few days. We must therefore pay a close attention to the number of efficiency evaluations needed.

We chose to test and compare three bounded local optimization methods that do not require the knowledge of the efficiency gradient: the so called CO\-BY\-LA \cite{powell_direct_1994,powell_direct_1998,powell_view_2007}, S\-L\-S\-Q\-P \cite{kraft_software_1988} and L-BFGS-B \cite{byrd_limited_1995,zhu_algorithm_1997,morales_remark_2011} methods.
The optimization work has been done using the SAGE \cite{sage} interface to the SciPy \cite{van_der_walt_numpy_2011,scipy} optimizers.

Each of the three methods separately took about twenty hours of computing time and yielded a maximum cell efficiency of $18.2\%$ for the following parameters: $\left(\X,\WF,\N,\ThickInGaN\right)=\left(0.56,6.3\mathrm{eV},6.5\times10^{16}\mathrm{cm^{-3}},0.86\mathrm{\mu m}\right)$.

As a check, we also used the three local methods using the mid-point of the parameters ranges as the starting points. Again, about twenty hours of computing time were necessary to yield the same result as above.

\section{Schottky Solar Cell Performances}

\subsection{Electrical Characteristics}
\label{electricalcharacteristics}

\begin{figure}
  \includegraphics[width=\linewidth]{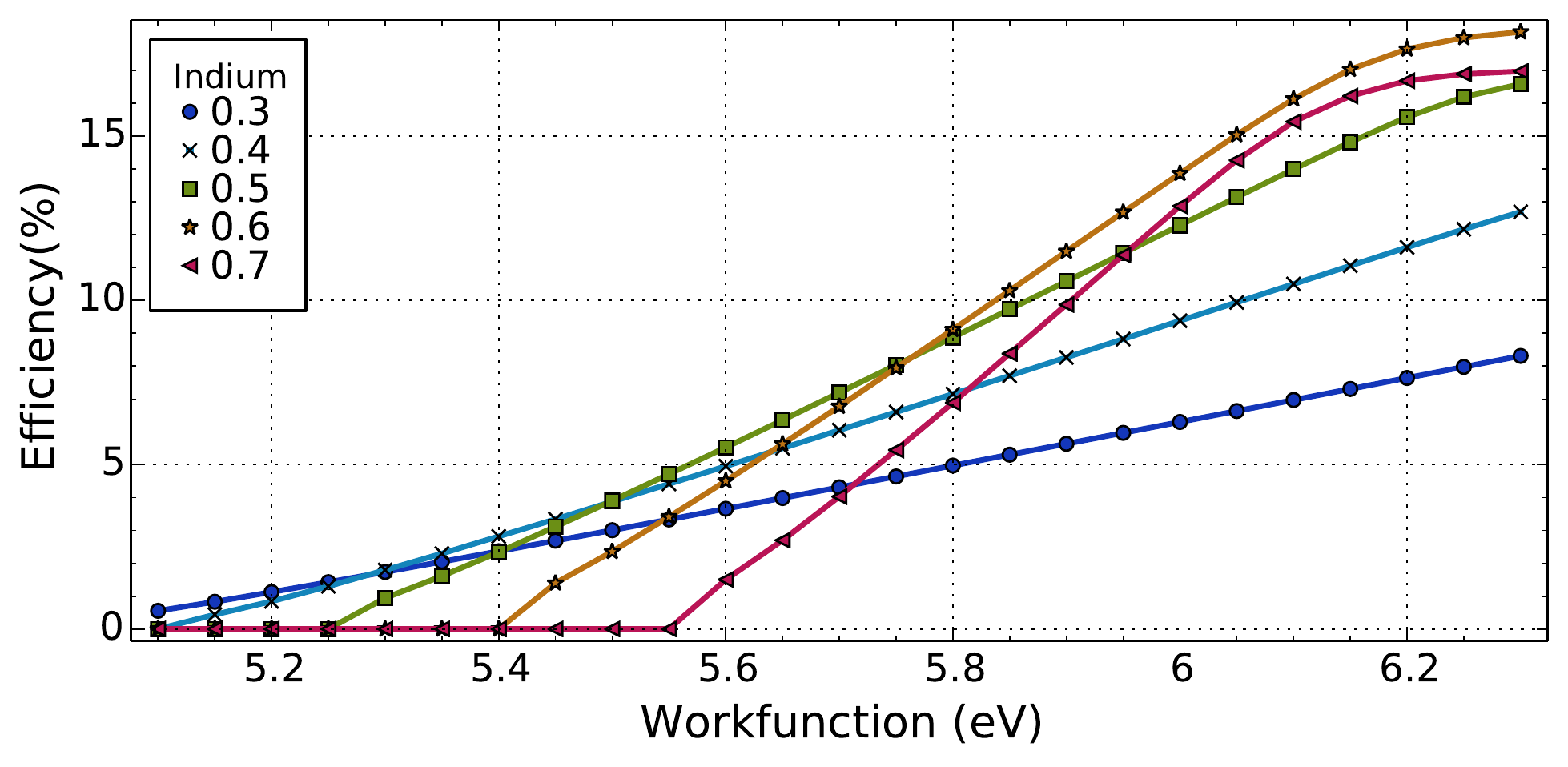}
  \caption{Efficiency as a function of the metal work function $\WF$ for different Indium compositions $\X$, for a doping of $\N=6.5\times10^{16}\mathrm{cm^{-3}}$ and an InGaN layer thickness of $\ThickInGaN=0.86\mathrm{\mu m}$.}
  \label{SchottkInGaN_rev_Fig2}
\end{figure}

\begin{figure}
  \includegraphics[width=\linewidth]{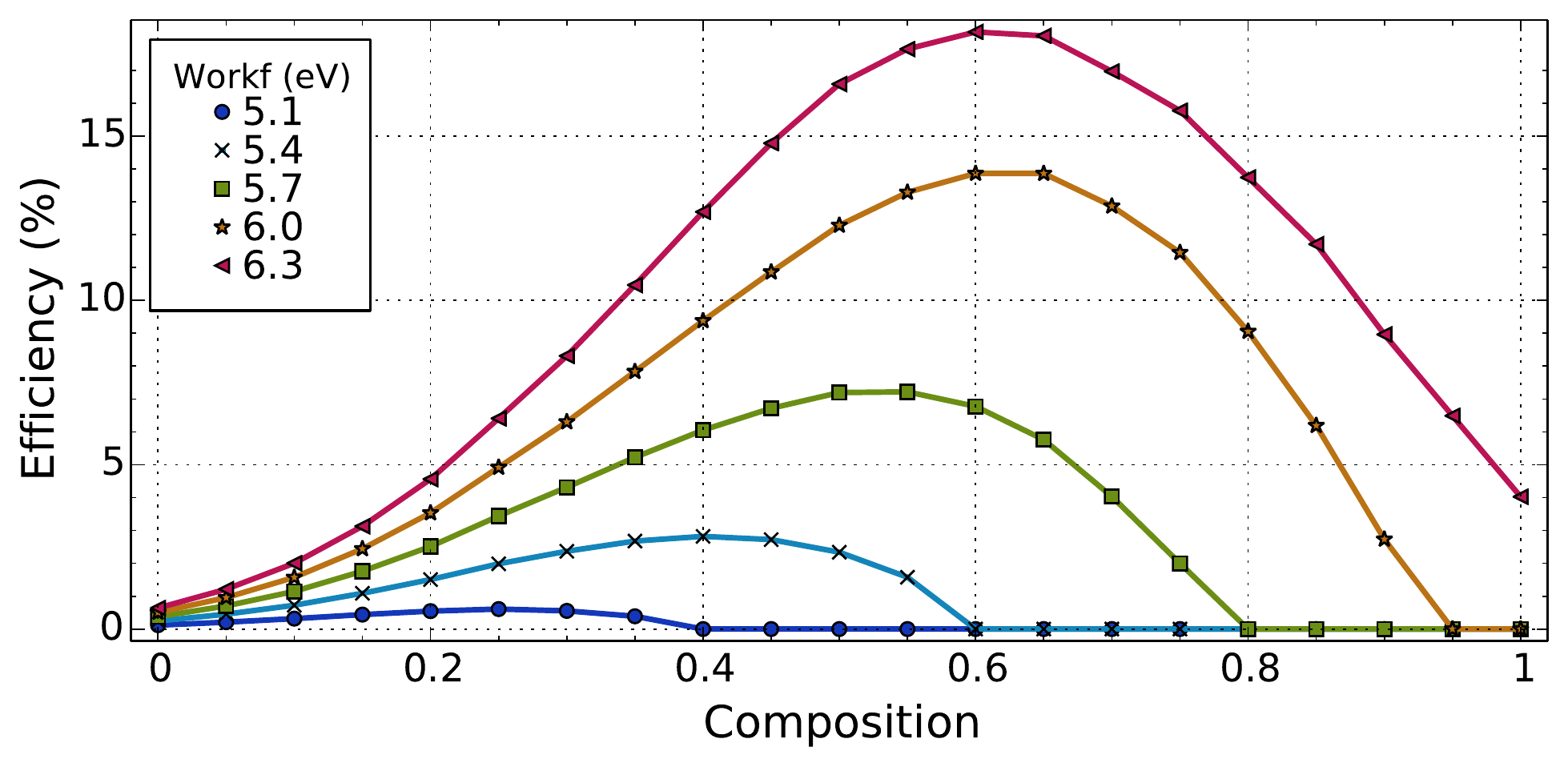}
  \caption{Efficiency as a function of the Indium compositions $\X$ for different metal work function $\WF$, for a doping of $\N=6.5\times10^{16}\mathrm{cm^{-3}}$ and an InGaN layer thickness of $\ThickInGaN=0.86\mathrm{\mu m}$.} 
  \label{SchottkInGaN_rev_Fig3}
\end{figure}

Fig. \ref{SchottkInGaN_rev_Fig2} shows the photovoltaic efficiency with respect to the metal work function for different Indium compositions and, accordingly, Fig. \ref{SchottkInGaN_rev_Fig3} shows this efficiency with respect to the Indium composition and for different metal work function.

\begin{figure}
  \includegraphics[width=\linewidth]{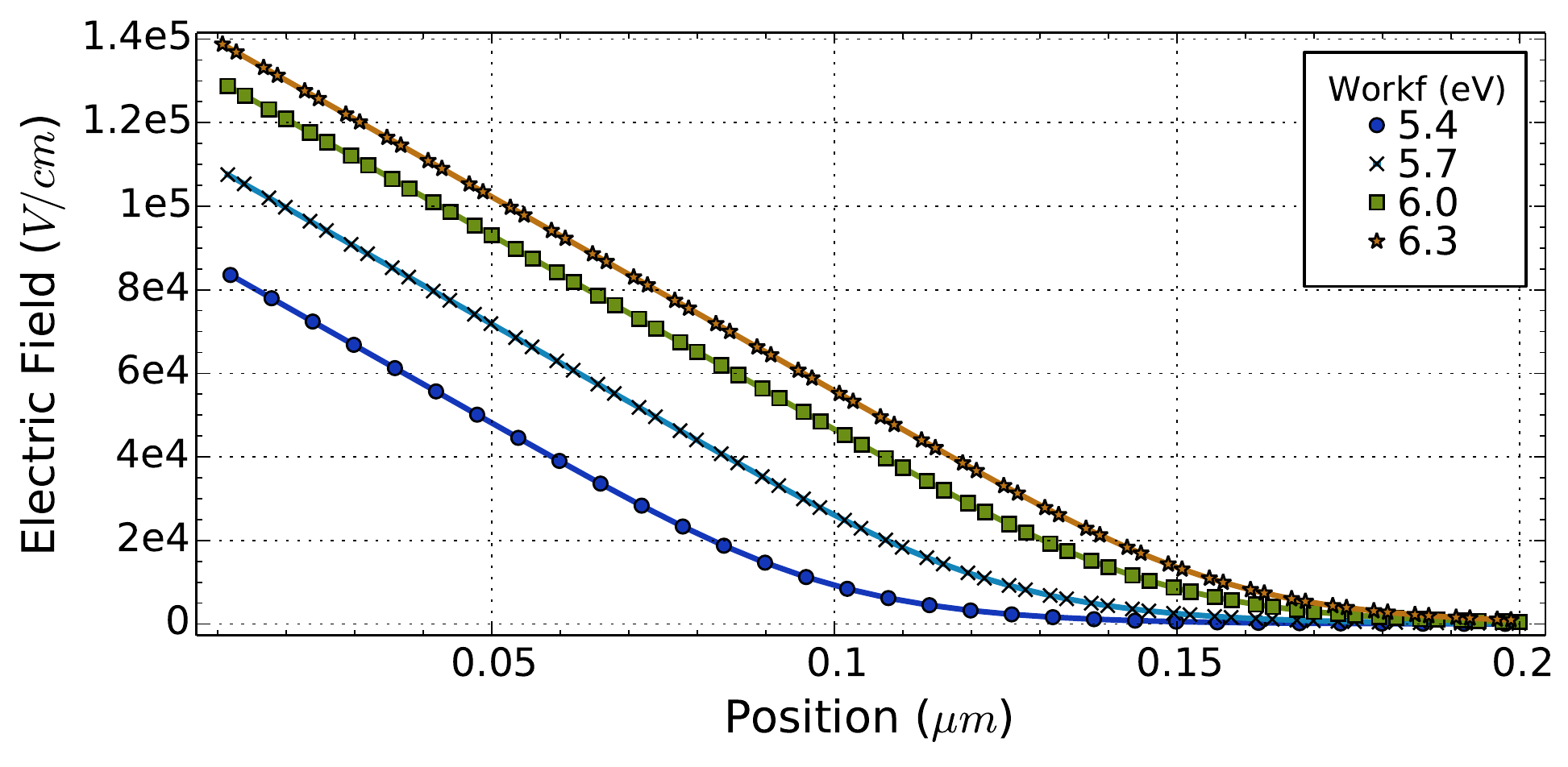}
  \caption{Electric field as a function of the distance from the device surface for different metal work function $\WF$, for a doping of $\N=6.5\times10^{16}\mathrm{cm^{-3}}$, an Indium composition of $\X=0.56$ and an InGaN layer thickness of $\ThickInGaN=0.86\mathrm{\mu m}$.} 
  \label{SchottkInGaN_rev_Fig4}
\end{figure}

\begin{figure}
  \includegraphics[width=\linewidth]{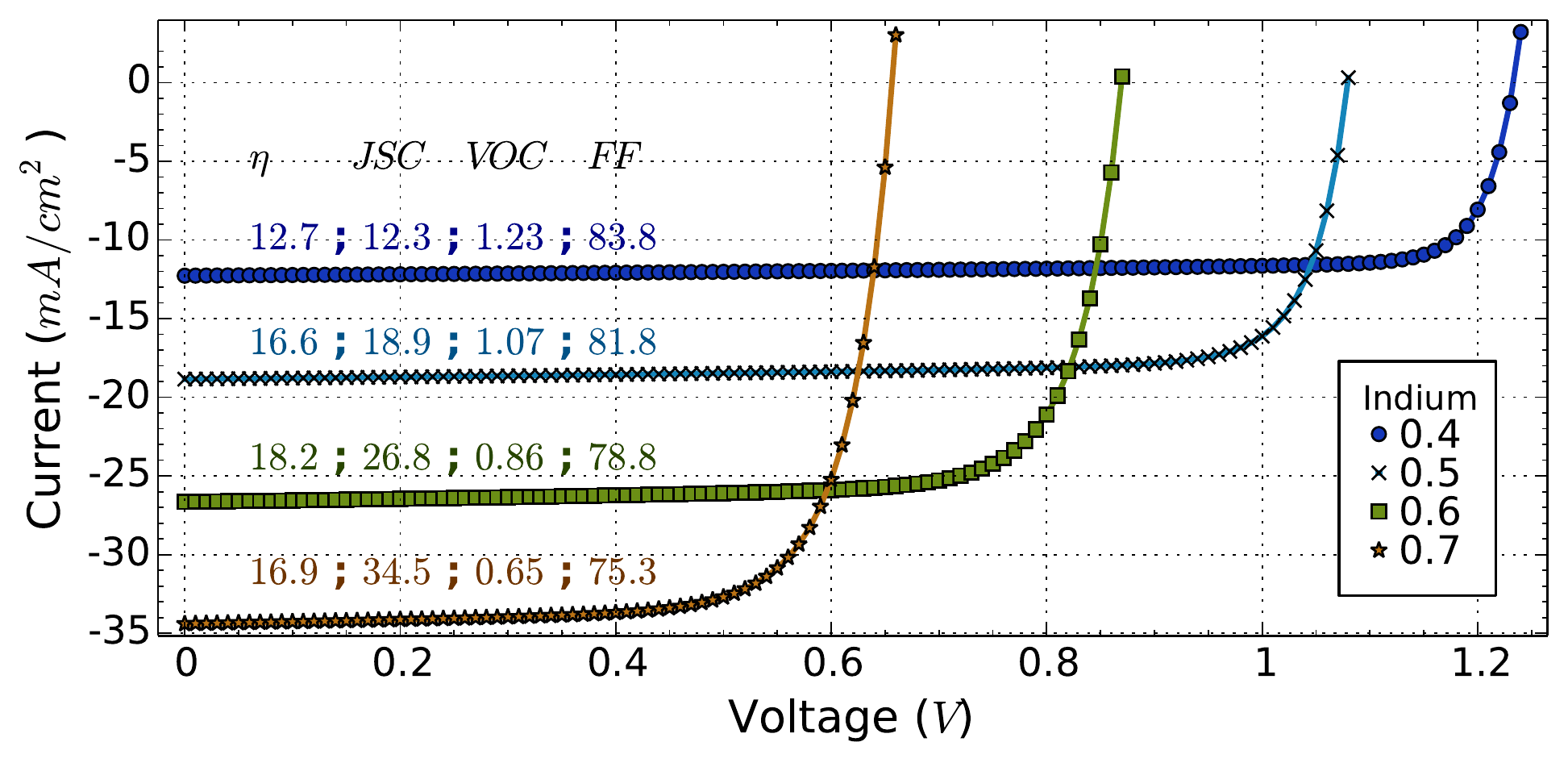}
  \caption{Current - Voltage characteristic for for different Indium compositions $\X$, for a metal work function $\WF=6.3\mathrm{eV}$, a doping of $\N=6.5\times10^{16}\mathrm{cm^{-3}}$ and an InGaN layer thickness of $\ThickInGaN=0.86\mathrm{\mu m}$. $\eta$, JSC, VOC and FF are respectively abbreviations for Efficiency ($\%$), Short-Circuit current density ($\mathrm{mA}/\mathrm{cm^{2}}$), Open-Circuit voltage (V) and Fill Factor. } 
  \label{SchottkInGaN_rev_Fig5}
\end{figure}

\begin{figure}
  \includegraphics[width=\linewidth]{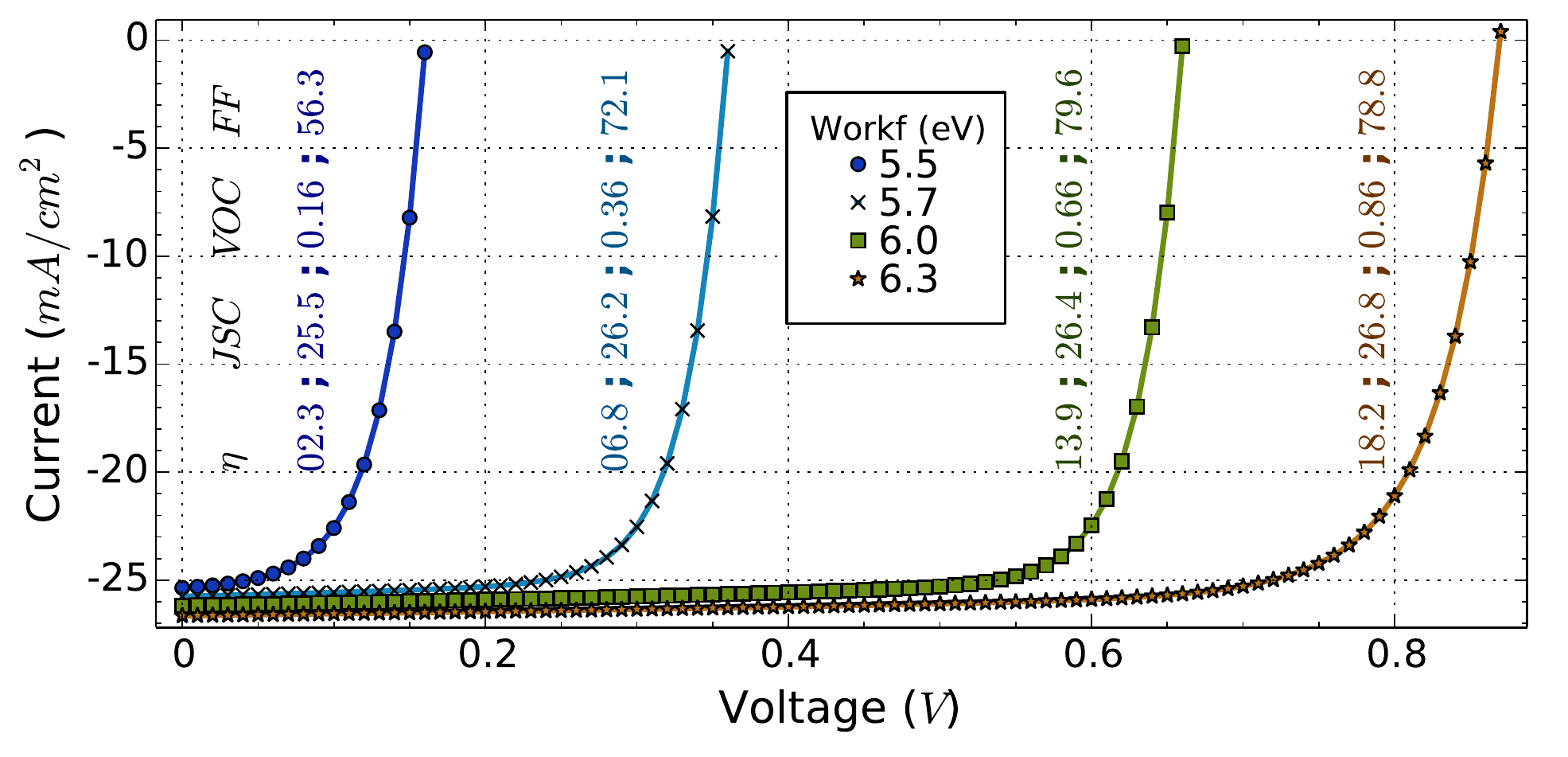}
  \caption{Current - Voltage characteristic for different work functions, for an Indium composition of $\X=0.56$, a doping of $\N=6.5\times10^{16}\mathrm{cm^{-3}}$ and an InGaN layer thickness of $\ThickInGaN=0.86\mathrm{\mu m}$. $\eta$, JSC, VOC and FF are respectively abbreviations for Efficiency ($\%$), Short-Circuit current density ($mA/\mathrm{cm^{2}}$), Open-Circuit voltage (V) and Fill Factor. } 
  \label{SchottkInGaN_rev_Fig6}
\end{figure}

Fig. \ref{SchottkInGaN_rev_Fig4} displays the electric field variation in the active layer and Figs. \ref{SchottkInGaN_rev_Fig5} \& \ref{SchottkInGaN_rev_Fig6} show the I-V characteristics with respect to the Indium composition and metal work function respectively.

Figs \ref{SchottkInGaN_rev_Fig2} \& \ref{SchottkInGaN_rev_Fig3} show that both the metal work function and the active layer composition highly impact the cell efficiency. Indeed, at a given Indium composition, the increase of the metal work function widens the space charge region and increases the electric field amplitude, as shown in Fig. \ref{SchottkInGaN_rev_Fig4}. All other things being equal, this increases the overall solar cell efficiency. This increase is mainly due to a higher open-circuit voltage and, to a lesser extent, to a better fill factor, as shown in Fig. \ref{SchottkInGaN_rev_Fig6}.

When the Indium composition is increased for a given metal work function, the whole set of the active layer parameters changes, including the  absorption spectrum and the band gap, this latter decreasing.
This decrease of the band gap and the linked change of the absorption spectrum enhances the solar light absorption in the visible range and thus the photovoltaic performances.

These better performances are due to the fact that the short-circuit current increases sharply, owing to a higher photogeneration and higher quantum efficiency. Meanwhile, the open-circuit voltage decreases as the metal/InGaN barrier height decreases.

When the Indium composition becomes higher than a certain limit, depending on the metal work function, the overall efficiency decreases and, ultimately, becomes non significant at a given Indium composition, as can be seen on Fig. \ref{SchottkInGaN_rev_Fig3}. The optimal Indium composition is fixed on the one hand by the matching between the solar spectrum and the InGaN absorption spectrum (and thus the band gap) and, on the other hand, by the quantum efficiency directly impacted by the metal/InGaN barrier height through the space charge region thickness and the electric field magnitude.

The resulting highest efficiency is $18.2\%$ obtained for an Indium composition of $0.56$ and a metal work function of $6.3 \mathrm{eV}$. The active layer thickness is $0.86\mathrm{\mu m}$. This efficiency is comparable to the efficiency of the well-established thin-films solar cells \cite{green_solar_2015} and could make this new technology highly competitive for the photovoltaic industry since the fabrication cost is expected to be relatively low. Indeed, its extremely simple one-semiconductor-layer makes it more cost effective to fabricate than current conventional multi-layer and/or he\-te\-ros\-truc\-tu\-re pho\-to\-vol\-ta\-ic cells \cite{cho_influence_2012}. In addition, its fabrication technology is comparable to that of commercially available InGaN LEDs, even though the Indium composition that is necessary for photovoltaic applications is still the goal of much research work \cite{fischer_highly_2013,yong_structural_2014}. However, should we allow a maximum composition of $40\%$, Fig. \ref{SchottkInGaN_rev_Fig2} shows that the efficiency would remain around $13\%$. 

Similarly, the same Fig. \ref{SchottkInGaN_rev_Fig2} shows that lowering the work function to $6.0 \mathrm{eV}$ does not impair the efficiency that much. Furthermore, the literature accounts for measurements of the work function of Platinum as high as $6.35 \mathrm{eV}$ \cite{chen_effect_2007,haridas_effect_2011,dubridge1928photoelectric}, close enough to the theoretical optimum, as seen on the same figure \ref{SchottkInGaN_rev_Fig2}.

However, some measurements in the literature report a lower work function for Platinum: $5.65 \mathrm{eV}$ can for instance be found in ref. \cite{rotermund1990investigation}, $5.93 \mathrm{eV}$ in ref. \cite{book2008work} and $6.10 \mathrm{eV}$ in ref. \cite{derry1989work}\footnote{Are cited here only the most cited values of the Platinum work function.}.
We thus performed a full optimization by keeping only the work function constant and equal to each of these widely cited experimental work function values. For a Platinum work function of $5.65 \mathrm{eV}$ (as reported, e.g., in \cite{rotermund1990investigation}), the efficiency decreases to $\approx 7.7\%$ obtained for layer thickness of $1.68 \mathrm{\mu m}$, layer doping of $3.7\times10^{15}\mathrm{cm^{-3}}$ and Indium composition of $0.48$. The Indium composition decreases as expected to maximize the effective barrier height. Indeed, Lowering the metal work function should be accompanied with a lowering of the active layer affinity and thus a decrease in the Indium composition. But since the Indium composition impacts all other parameters (e.g. absorption spectrum) the overall cell efficiency will change accordingly. For a Platinum work function of $5.93 \mathrm{eV}$ (as reported, e.g., in \cite{book2008work}), the obtained optimal efficiency is $\approx 13.9\%$ obtained for layer thickness of $1.65 \mathrm{\mu m}$, doping of $8\times10^{15}\mathrm{cm^{-3}}$ and Indium composition of $0.58$. For a Platinum work function of $6.10 \mathrm{eV}$ (as reported, e.g., by \cite{derry1989work}), an optimal efficiency of $\approx 17\%$ was obtained for layer thickness of $1.16 \mathrm{\mu m}$, doping of $1.54\times10^{16}\mathrm{cm^{-3}}$ and Indium composition of $0.57$. The here obtained value is very close to the optimal one ($18.2\%$).

The thickness needed is $0.86\mathrm{\mu m}$ to achieve the highest efficiency ($18.2\%$). Should the InGaN thickness be $0.3\mathrm{\mu m}$, the corresponding efficiency would be $14.5\%$. Should it be $0.2\mathrm{\mu m}$, the efficiency would only decrease to $12.5\%$, remaining comparable to the efficiency of standard thin-films solar cells \cite{green_solar_2015}. The proposed solar cell could then be realized with very thin InGaN layer, down to a $200\mathrm{nm}$ thickness.

This latter has at least two advantages. One of them is to drastically minimize the quantity of precursors needed to elaborate the active layer, usually using Metalorganic Chemical Vapour Deposition (MOCVD). The other advantage is the possibility to elaborate high quality pseudomorphic layers. 

Finally, the optimal doping of the n-layer is comparable to the residual concentration usually measured for this alloy, possibly removing the need for additional doping.

\subsection{Bandgap Bowing Impact}
\label{bowing}

The literature exhibits a quite large consensus around the $b=1.43$ value of the bowing parameter \cite{wu_band_2003,fabien_guidelines_2014} used in this work, for thicknesses comparable to ours and corresponding to relaxed layer in the whole composition range. However theoretical \emph{ab initio} calculations \cite{moses_band_2010} cast some doubt on the very validity of the unique quadratic approximation for the whole composition range. In other words, the bowing factor could depend on the Indium composition.

These uncertainties on the bowing parameter actual value for relaxed layers lead us to propose here the quantitative value of the optimized parameters of the Schottky solar cell for another value of the bowing factor, $b=3.0 \mathrm{eV}$, as suggested by \emph{ab initio} calculations made by \cite{moses_band_2010}.

For this bowing factor value of $b=3.0 \mathrm{eV}$, the optimum cell efficiency is $17.8\%$ for the following parameters : $(\X,$ $\WF,$ $\N,$ $\ThickInGaN)=$ $\left(0.44,6.3\mathrm{eV},8.81\times10^{16}\mathrm{cm^{-3}},0.89\mathrm{\mu m}\right)$. It is only slightly lower than the value of $18.2\%$ that has been calculated previously for $b=1.43 \mathrm{eV}$.
Furthermore, the optimal composition value $x$ that is retrieved is lower than previously, as could be expected to keep the optimal band gap energy. The other parameters are similar.

\subsection{Interfacial Layer Effect}
\label{interfaciallayer}

The contact metal to III-Nitride layer is known to possibly exhibit thin interfacial layer which can be due to particle irradiation, thermal treatment such as annealing, diffusion, crystal polarization or surface defects \cite{jang2006schottky}. This can be modeled as the creation of a thin interfacial highly doped layer with a typical doping concentration around $10^{19}\mathrm{cm}^{-3}$ when intentionally induced by high energy particle irradiation \cite{li2005fermi} and around $5\times10^{18}\mathrm{cm}^{-3}$ when due to native defects \cite{mamor2009interface}. We thus added in our model a 5 nm $n^+$ layer in between the n-doped layer and the metal and made calculations for the interfacial layer doping density varying from $1\times10^{17}\mathrm{cm}^{-3}$ to $1\times10^{20}\mathrm{cm}^{-3}$. This last maximum value is one order of magnitude higher than the value intentionally induced by particle irradiation in \cite{li2005fermi}. We have however included this extreme value in the simulations to give an insight on what would happen then.

\begin{figure}
  \includegraphics[width=\linewidth]{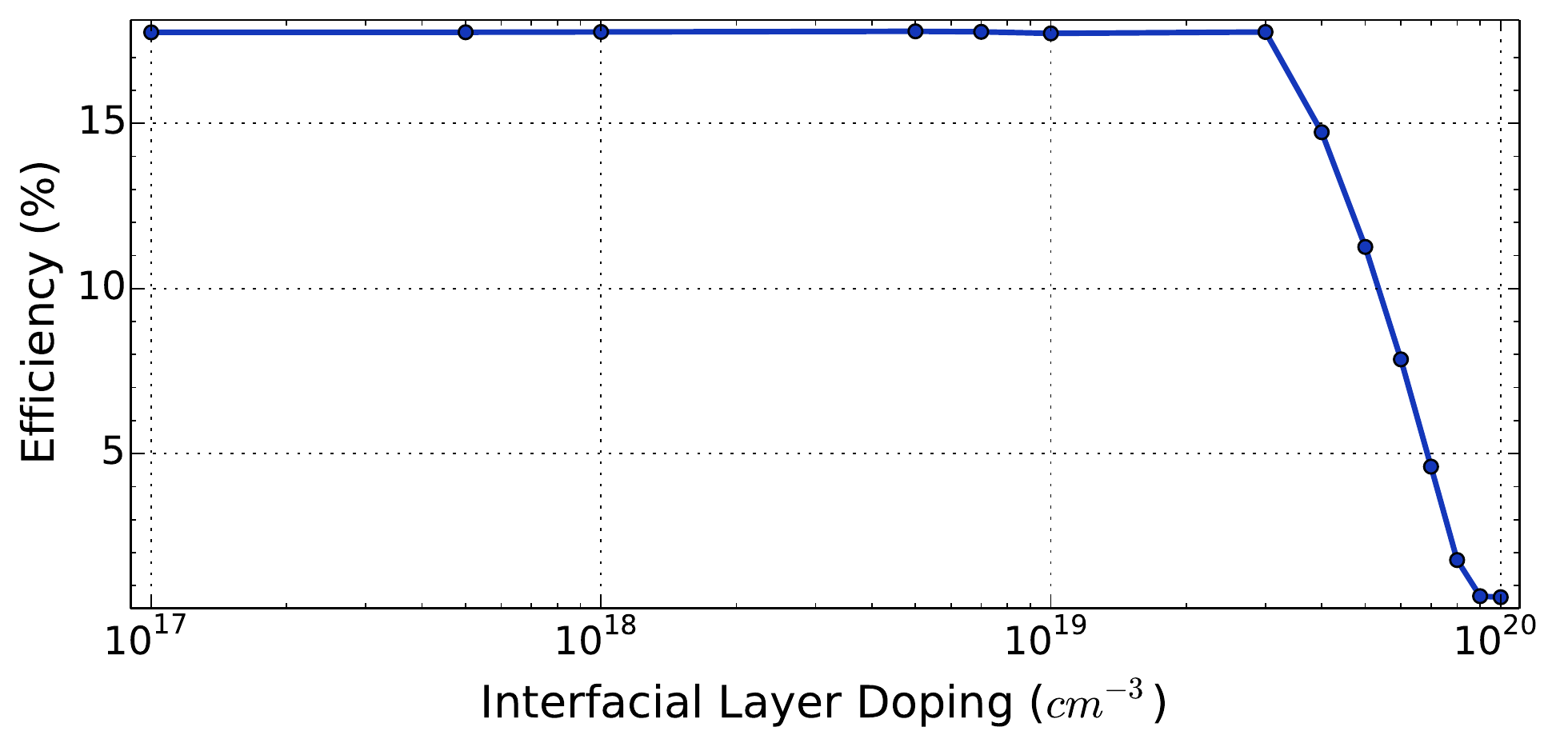}
  \caption{Efficiency as a function of the interfacial layer doping concentration, for a metal work function $\WF=6.3\mathrm{eV}$, an Indium composition of $\X=0.56$, a doping of $\N=6.5\times10^{16}\mathrm{cm^{-3}}$ and an InGaN layer thickness of $\ThickInGaN=0.86\mathrm{\mu m}$.}
  \label{SchottkInGaN_rev_Fig7}
\end{figure}

Fig. \ref{SchottkInGaN_rev_Fig7} shows the efficiency variation with the interfacial layer doping. This efficiency, obtained for the optimal parameters $\left(\X,\WF,\N,\ThickInGaN\right)$ $=\left(0.56,6.3\mathrm{eV},6.5\times10^{16}\mathrm{cm^{-3}},0.86\mathrm{\mu m}\right)$, remains almost unchanged for dopings below $4\times10^{19}\mathrm{cm}^{-3}$ and decreases down to $11.25\%$ for an extreme doping value of $5 \times10^{19}\mathrm{cm}^{-3}$. For the very high $1\times10^{20}\mathrm{cm}^{-3}$ doping concentration, the efficiency dramatically decreases to less than $1.0\%$.
The threshold effect, observed at the value of $4 \times10^{19}\mathrm{cm}^{-3}$, is induced by the $n^+-n$ junction contribution, at these very high interfacial doping levels. It indeed impacts the overall solar cell I-V characteristic, resulting in a rapid decrease of the fill factor which is brought down from $79\%$ to $54\%$ when the $n^+$ doping concentration increases from $1\times10^{19}\mathrm{cm}^{-3}$ to $5\times10^{19}\mathrm{cm}^{-3}$ as shown in table \ref{tabinterfacial}.

\begin{table}
\begin{center}
\begin{tabular}{|c|c|c|c|}
\hline
$\Unit{N^+}{cm^{-3}}$ & $\E{1}{19}$ & $\E{3}{19}$ & $\E{5}{19}$ \\ \hline
$\Unit{JSC}{mA/cm^{2}}$ & 23.34 & 23.07 & 22.63 \\ \hline
$\Unit{VOC}{V}$ & 0.96 & 0.97 & 0.92 \\ \hline
$\Unit{FF}{\%}$ & 79.2 & 79.0 & 54.04 \\ \hline
$\Unit{\eta}{\%}$ & 17.74 & 17.67 & 11.25 \\ \hline
\end{tabular}
\end{center}

\caption{Short-Circuit Current ${JSC}$, Open-Circuit Voltage ${VOC}$, Fill Factor FF and Efficiency ${\eta}$ for some interfacial layer doping concentrations, for a metal work function $\WF=6.3\mathrm{eV}$, an Indium composition of $\X=0.56$, a doping of $\N=6.5\times10^{16}\mathrm{cm^{-3}}$ and an InGaN layer thickness of $\ThickInGaN=0.86\mathrm{\mu m}$.\label{tabinterfacial}}
\end{table}

\section{Conclusion}

We have studied, by using mathematical optimization methods, the new InGaN Schottky solar cell. We have applied sophisticated optimization algorithms, for the first time to our knowledge, to the study of InGaN based solar cells. This has allowed to demonstrate the usefulness of such an approach to the study of novel solar cells, compared to the classical parametric analysis. Using the most realistic physical parameters, obtained from previously published experimental papers, we obtained a maximum conversion efficiency of $18.2\%$ for the InGaN Schottky solar cell. The demonstrated figure of merit of the simulated InGaN Schottky solar cell, the realistic parameters set and the expected elaboration simplicity make the InGaN Schottky solar cell an attractive alternative to the usual InGaN p-n and p-i-n junction solar cells.





\section*{References}

\bibliographystyle{elsarticle-num} 
\bibliography{SchottkInGaN_rev_Ref}

%

\end{document}